\documentclass[preprint,journal]{rmaa}
\newcommand\Mdot{\dot{M}}
\newcommand\degr{^\textrm{\scriptsize o}}

\suppressfulladdresses 

\usepackage{paralist}

\usepackage{psfrag,color}

\SetYear{2006}
\SetConfTitle{RevMexAA}

\title{STATISTICAL ANALYSIS OF MOLECULAR LINE EMISSION FROM T TAURI DISK MODELS} 

\author{
  Itziar de Gregorio-Monsalvo,\altaffilmark{1} 
  Paola D'Alessio,\altaffilmark{2}
  and Jos\'e F. G\'omez\altaffilmark{3}}

\altaffiltext{1}{European Southern Observatory, Alonso de C\'ordova 3107, Vitacura, Santiago, Chile (idegrego@eso.org).}
\altaffiltext{2}{Centro de Radioastronom\'\i{}a y Astrof\'\i{}sica,
  Universidad Nacional Aut\'onoma de M\'exico, Campus Morelia, Apartado Postal 3--72, 58090, Morelia,
  Michoac\'an, M\'exico (p.dalessio@astrosmo.unam.mx).}
\altaffiltext{3}{Instituto de Astrof\'{\i}sica de Andaluc\'{\i}a (CSIC), Apartado 3004, E-18080 Granada, Spain (jfg@iaa.es).}

\shortauthor{de Gregorio-Monsalvo, D'Alessio, \& G\'omez}
\shorttitle{MODELING MOLECULAR EMISSION LINES FROM T TAURI DISKS}

\listofauthors{I. de Gregorio-Monsalvo, P. D'Alessio, \& J. F. G\'omez}
\indexauthor{de Gregorio-Monsalvo, I.}
\indexauthor{D'Alessio, P.}
\indexauthor{G\'omez, J. F.}

\abstract{In this work we model the expected emission from the molecular line  C$^{17}$O(J=3 $\rightarrow$2) in
protoplanetary disks, modifying different physical parameters to
obtain distinctive observational signatures. Our aim is to determine the kind of
observations that will allow us to extract information about the
physical parameters of disks. With this purpose we perform a
statistical analysis of principal components and a multiple linear
correlation on our set of results from the models. We also present prospects for
future molecular line observations of protoplanetary disks using SMA
and ALMA.
}

\resumen{En este trabajo modelamos la emisi\'on esperada de la l\'{\i}nea molecular C$^{17}$O(J=3 $\rightarrow$2) en discos protoplanetarios, modificando diferentes parametros f\'{\i}sicos para obtener distintas caracter\'{\i}sticas observacionales. Nuestra meta es determinar la clase de observaciones que nos permitir\'an extraer informaci\'on sobre los par\'ametros f\'{\i}sicos de los discos.  Con este prop\'osito realizamos un an\'alisis estad\'{\i}stico de componentes principales y una correlaci\'on lineal m\'ultiple en el conjunto de resultados obtenidos a partir de los modelos. Adem\'as, presentamos un estudio sobre futuras observaciones de l\'{\i}nea molecular en discos protoplanetarios usando SMA y ALMA.}

\addkeyword{ISM: molecules}
\addkeyword{method: statistical}
\addkeyword{radiative transfer}
\addkeyword{Stars: planetary systems: protoplanetary disks}

\begin{document}
\maketitle

\section{Introduction}
\label{PPD:int}
The mass of disks around young stars is $\sim 99\%$ gas and only 1\% dust. 
However, since the dust opacity is large compared to the gas opacity 
in a wider range of wavelengths, the dust component dominates the absorption 
and reprocessing of stellar radiation 
and the emergent spectral energy distributions (SEDs) of disks 
around Classical T Tauri (CTTS) and Herbig Ae stars (HAe).
Most of the models constructed to explain observed 
SEDs have taken into account only the dust component in the calculation 
of the disk temperature \citep{Chi97, Dal98, DAl99, DAl01}. 
These models have been successful to explain the observed SEDs.
However, SEDs are not sensitive to kinematical information, details of the
radial and vertical temperature distribution, the chemistry  of the gaseous component
\citep{Aik96,Aik97,Aik99,VanZad01,Aik06}, 
the possibility that the gas in the upper layers is hotter than the
dust in lower layers
by absorption of UV and X-rays (i.e., \citealt{Gla04,Jon04,Kam04})
 and/or photo-electrical effect (i.e., \citealt{Nom05}), etc.

The study of molecular line emission from disks around young stars is 
an important tool to infer physical characteristics of disks
(e.g., \citealt{Dar03,Car04,Pie05,Qi06,Ram06,Dut07}). 
An advantage of a spectral line
is that emission at different frequencies/velocities might be probing different
disk regions, making lines an important test for disk models. However,
this is also a disadvantage in some sense,  
 since one observes intensities convolved with the beam of the
 telescope and with a finite 
spectral resolution, and therefore the information of the different regions 
is mixed up in a complex way.  Thus, the analysis and relationship between 
observations and model properties might become very complicated and difficult to disentangle.
There are some previous works that compare molecular line emission from
  protoplanetary or circumbinary disks with specific models of such an
  emission (e.g., \citealt{Koe93,Gui98,Qi03,Qi04}) and, in general,
  agreement between model and observations is fairly good, at least in the
  general appearance of the maps. However, given the great deal of physical
  parameters involved in the resulting molecular line emission, it is not
  straightforward to determine those parameters from a particular observation
  just by fitting an emission model. 

Our main aim in this paper is to identify a set of observational characteristics
that give most information on the physical properties of the
disk. Such observational characteristics should then be given the heaviest
weight in a fit between observations and model aiming to determine physical
parameters in a disk. 

This paper is structured as follows: in \S\ref{PPD:gen} we show the outline of this work. In \S\ref{PPD:str} we describe the
assumptions to calculate the disk structure models and the initial
input parameters. In \S\ref{PPD:rad}  we explain the radiative transfer
calculation and 
we discuss the selection of the C$^{17}$O (J=$3\rightarrow 2$) as  the molecular line
to make our study. In \S\ref{PPD:gri} we outline the network of models and the general trends of the line 
emission maps.
In \S\ref{PPD:sta} we describe the statistical study that identify the
 set of observational characteristics
that give most information on the physical properties of the
disk, and we comment the results derived in \S\ref{PPD:res}. 
In \S\ref{PPD:com} we show the comments and prospects for
these studies.
Finally, in \S\ref{PPD:dec} we perform a
study of the detectability of our modeled disks with SMA and ALMA and we summarize the conclusions in \S\ref{PPD:con}. 

\section{General considerations}
\label{PPD:gen}

In order to achieve the goal described in the introduction, we have developed a set of molecular line emission models calculated for various mass accretion rates, radius, viscosities and maximum dust grain radius distributions. 
The small size scale of a protoplanetary disk ($\sim$100 AU) and
their low temperatures ($\sim$ 100 K; \citealt{Bec90,Miy95,Bec00})
require observations with high sensitivity and subarsecond high-angular
resolution, since 100 AU subtends  $0\farcs 7$ at 140 pc (the distance
to the Taurus cloud). This makes interferometric observations
necessary to carry out this kind of studies. 
With the intention of
reproducing a real interferometric observation of a protoplanetary disk with
different physical parameters, we integrated the radiative transfer equation and convolved each model with a beam
of $0\farcs 4$, as a compromise between resolution and
sensitivity. From each resultant map, we have measured different observational signatures, as if they were data from a real interferometric observation.
Finally, in order to obtain the best combinations of such
observational parameters that yield more information about the
physical characteristics of disks, we have undertook a novel
statistical approach to link observational properties of the expected
molecular line emission with the underlying physical properties of the
disk, by means of a principal component and multiple linear
correlation analysis. We show that this is a promising
type of analysis to prepare the observations with the new generation
of millimeter and submillimeter interferometers.


In this study it is important to choose an appropriated molecular
transition  sensitive to the physical parameters in which we may be
interested. To carry out our study
we have selected the C$^{17}$O(J=$3\rightarrow 2$) transition at 337 GHz.
This line is  a high excitation transition of a CO isotope with very  
low abundance, which makes it less susceptible to be affected by absorption 
and/or the emission from the surrounding cloud material. 
 This transition is also a suitable candidate to be 
observed in protoplanetary disks using the Submillimeter Array (SMA) and the
Atacama Large Millimeter Array (ALMA), as shown by
\citet{Gom00}. 

\section{Disk structure models}
\label{PPD:str}
\subsection{Assumptions}

We base our calculations of molecular line emission on structure models
of accretion disks irradiated by the central star,  
which have been previously used  to explain different observations 
of classical T Tauri Stars.
The assumptions and calculation method of such models
are described in \citet{Dal98, DAl99, DAl01}. In summary, 
the disk is assumed to be in steady state, with a constant 
mass accretion rate $\Mdot$ and an $\alpha-$viscosity 
\citep{Sha73}, with a constant value of the viscosity parameter $\alpha$.
The disk is in vertical hydrostatic equilibrium in 
the gravitational potential well of the star, and we neglect the 
disk self-gravity.
We assume that gas and dust are
thermally coupled, having the same temperature everywhere. This 
dust/gas temperature enters in the calculation of the 
disk volumetric density distribution through the 
integration of the hydrostatic equilibrium equation.
The main heating mechanisms considered are viscous dissipation 
and stellar irradiation.  The viscous dissipation  is important in heating the 
inner regions (close to the star and close to the midplane), the  
direct stellar irradiation heats the disk atmosphere, 
and the stellar radiation scattered and reprocessed by the disk 
upper layers heats the whole vertical structure.
The transfer of radiation  through the disk is calculated taking into 
account that the dust scatters and absorbs stellar and disk radiation, 
implying that the temperature structure depends on the dust properties.
The viscous irradiated disk models used in the present study show 
the temperature inversion previously found by \citet{Cal91,Cal92}, 
i.e., at the outer disk, $R \gtrsim $ 10 AU, the upper layers are hotter 
than the disk midplane, because they are heated by direct stellar irradiation 
(see also \citealt{Chi97} and \citealt{Dal98}). 

The dust opacity is calculated using the Mie theory for compact 
spherical grains. 
We consider a distribution of sizes given by 
$n(a)=n_{0}a^{-p}$, where $a$ is the radius of the grains, 
$n_{0}$ is a normalization constant, and $p$ is a free parameter. In this 
work, we have adopted p~=~3.5 from \citet{Dra84} and the model of dust 
composition proposed by \citet{Pol94} with the variations introduced
 by \citet{DAl01}. To account for the possibility of dust growth, 
we adopt different values of maximum grain sizes.      
Dust grains of different sizes have different continuum opacity 
at mm wavelengths, affecting the molecular line emission in different 
ways. 
The existence of bigger grains in disks than in the interstellar medium 
was proposed to explain the observed slope 
of the continuum SED at millimeter wavelengths   
\citep{Bec91,Miy95}. Large grains could be depleted from higher layers
of the disk, but could be well mixed with gas below a few gas scale heights. 
For simplicity, the disk models adopted here (from \citealt{DAl01}), assume that gas 
and dust are well mixed. This seems a reasonable assumption if the emission from
the molecular line arises mainly from areas closer to the midplane than to the upper layers.  This point will
be discussed later (subsection~\ref{PPD:rad_line}).

\subsection{Input parameters}

For the present study we have considered the following input parameters:
maximum disk radius ($R_d$), maximum radius of dust grains ($a_{max}$), disk mass accretion rate ($\Mdot$), and viscosity parameter ($\alpha$) (see values in Table~\ref{tb:PPD_phys-param}).

\begin{table}[!h]
\setlength{\tabnotewidth}{0.6\columnwidth}
  \tablecols{4}
{\small
\begin{center}
\caption[Initial physical parameters]
{Initial physical parameters}
\label{tb:PPD_phys-param}
\vspace{2mm}
  \begin{tabular}{cccl}
    \toprule \toprule
\multicolumn{1}{c}{$R_{d}$\tabnotemark{a}}&
\multicolumn{1}{c}{$a_{max}$\tabnotemark{b}}&
\multicolumn{1}{c}{$\dot{M}$\tabnotemark{c}} &
\multicolumn{1}{c}{$\alpha$\tabnotemark{d}}\\
\multicolumn{1}{c}{(AU)}  & 
\multicolumn{1}{c}{($\mu$m)}&
\multicolumn{1}{c}{(M$\sun$/year)}\\
\midrule
50  & 1        & 10$^{-9}$   & 0.001\\
100 & 10       & 3 10$^{-8}$ & 0.005\\
150 & 10$^{2}$ & 10$^{-7}$   & 0.01 \\
    & 10$^{3}$ &             & 0.02 \\
    & 10$^{4}$ &             & 0.05 \\
    & 10$^{5}$ &             &      \\
\toprule \toprule

 \tabnotetext{a}{Disk radius.}
 \tabnotetext{b}{Maximum radius of dust grains.}
 \tabnotetext{c}{Mass accretion rate.}
 \tabnotetext{d}{Viscosity parameter.}
\end{tabular}
\end{center}
}
\end{table}

We have adopted typical parameters of a T Tauri star from \citet{Gul98}, 
i.e., $M_{*}$ = 0.5 M$_{\sun}$, $R_{*}$ = 2 R$_{\sun}$, and $T_{*}$ = 4 000 K 
for all the models. The disks are assumed to be 
at 140 pc, the distance of the Taurus molecular cloud 
\citep{Ken94}, with 
a typical inclination angle i = 60$\degr$. 
It is important to mention that each disk structure is self-consistently 
calculated given these input parameters.
This means that the whole disk structure is affected by all the parameters, 
consequently affecting the line properties.
This might complicate the analysis of the resulting line properties,
but we think this gives a more realistic description of the interplay between 
the different variables.

\section{Radiative transfer and molecular line emission}
\label{PPD:rad}

The model of the disk structure provides a detailed density and temperature 
distribution through the disk as a function of the height and the distance 
to the disk center. 
To derive the line intensity for a given molecular transition, we must resolve
 the transfer equation. We have used the same assumptions and formalism that
 \citet{Gom00}, and we summarize them briefly here. 
We assume local thermal equilibrium for the population of the molecular 
energy levels and we consider thermal line profiles. We divide the disk in a 
grid of cells considering isovelocity lines and their perpendicular
lines (see Appendix in 
\citealt{Gom00}). We integrate the transfer equation  
$dI_{\nu}/ds = \kappa_{\nu}\rho(S_{\nu} - I_{\nu})$ through the line of sight
 at the center of each cell, where $I_{\nu}$ is the intensity, $s$ is the length along the line of sight, $\kappa_{\nu}$ 
is the absorption coefficient that consider the contributions from the line and
 continuum, i.e., $\kappa_{\nu}=\kappa_{l}+\kappa_{c}$, $\rho$ is the mass 
density of the gas, and $S_{\nu}$ the source function. The coefficient
$\kappa_{c}$ is dominated by dust and we consider pure absorption opacity of 
each kind of dust grains size. 



In order to derive the flux density, we convolved all our models with a beam of
 0$\farcs$4 of HPBW, as a compromise between resolution and sensitivity.
Finally we subtract the continuum emission to isolate the flux density of 
the molecular line, obtaining a set of model results that reproduce a real 
observation of a protoplanetary disk with different physical parameters.

\subsection{Selection of the emission line transition}
\label{PPD:rad_line}
It is very common that young stellar objects like T Tauri stars are still 
embedded in the material of the parental cloud. The envelope that
surrounds the disk-star system is composed by cold gas and dust that
could hide the disk emission. 

On one hand, the surrounding material could absorb the emission from 
the hotter, innermost part of the disk structure. This problem can be solved
by selecting a molecular transition whose energy is high enough to trace the
hotter gas from the disk, while few molecules in the colder envelope are in
the states involved in this transition, thus avoiding line absorption at the
envelope. However, the frequency of the transition must not be too high,
because otherwise the dust in the envelope would become optically thick.
On the other hand, the molecular line emission from the whole cloud
could hide the deeper emission from the disk, as it is usually the case for the more abundant
CO isotopes. To make sure that the observed emission comes only from the
disk structure, we must select a molecular species of low abundance for which
the envelope is optically thin and therefore its emission would be negligible.

In order to fulfill these conditions, we have chosen the (J=$3\rightarrow 2$)
transition of the C$^{17}$O molecule at 337 GHz. Its molecular
abundance relative to H$_{2}$ in molecular cores is low, 5.0$\times$10$^{-8}$
\citep{WhiS95} and the frequency transition at 337 
GHz is still low enough to avoid being much absorbed by the dusty envelope.

In a recent work, \citet{Dar03} have studied the disk vertical 
temperature structure using different isotopes of CO. 
Being characterized by different opacities, different lines  trace the gas at 
distinct depths. \citet{Dar03} find a good agreement between 
the inferred temperature for each transition/isotope and  the temperature 
where $\tau\sim 1$ in irradiated (dusty) disk models.
Following \citet{Dar03}, we have studied the formation regions of 
 molecular lines from different isotopes of CO (CO, $^{13}$CO, C$^{18}$O,
and C$^{17}$O) in our set of 
disk models, resulting that the  
C$^{17}$O is formed closer to the midplane than the rest of the
isotopes. This fact makes the LTE assumption acceptable, since the
region that mainly contribute to the line emission shows higher densities than the critical density
of the selected transition ($\simeq 5\times 10^4$ cm$^{-3}$).

In addition, given the high gas densities in the disk, we expect that
the abundance of this isotope, would be 
less affected by photodisociation produced by the incident
radiation, since it would be shielded against it, specially in the
deepest layers. Moreover, 
theoretical models of the evolution of molecular abundances in
protoplanetary disks, predict depletion of CO from the gas phase for
temperatures below 20 K \citep{Aik96} and therefore, depletion is
probably not significant for the chosen molecule and for the disk
radii we are considering here ($< 150$ AU). Thus, for simplicity, we adopt a
constant abundance for C$^{17}$O relative to H$_{2}$, given the molecular core
value \citep{WhiS95}.


\section{Results}
\label{PPD:gri}

\begin{figure}[!h]
\includegraphics[width=\columnwidth]{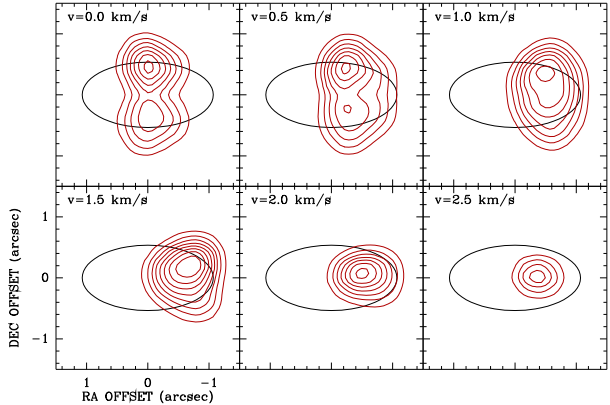}
\caption
{Emission maps at different velocities for a disk with radius
  = 150 AU, a$_{max}$ = 10 $\mu$m, i = $60\degr$, $\alpha$ = 0.01 and mass accretion rate =
  $10^{-7} M_{\odot}$ year$^{-1}$. Maps have been convolved with a
  0$\rlap.{''}$4 beam. The lowest contour and the increment
  step are 20 mJy beam$^{-1}$. The ellipse traces the outer edge of the disk.} 
\label{fig:PPD_disk}
\end{figure}

We have calculated the expected emission in the  C$^{17}$O(J=$3\rightarrow 2$) transition for a typical T Tauri star with a disk inclination angle
 of 60$\degr$, and for all possible combinations 
  of the physical parameters shown on
  Table~\ref{tb:PPD_phys-param}. We have 241 different models for
  which 
 we have integrated the radiative transfer equation at 12 different velocities, 
from -2.5 to 2.5 km s$^{-1}$ at steps of 0.5 km s$^{-1}$. The results are 
maps like those shown in Fig \ref{fig:PPD_disk}, in which we only show the positive velocities,
 since the maps are similar and almost symmetrical in negative velocities to the ones shown here
 (there are slight differences though, due to the hyperfine structure of
 C$^{17}$O transition). 
From each map we have measured the following observational signatures
as if they were data from a real 
interferometric observation: intensity of the principal (the more distant from the observer, to the north in our maps) and 
secondary (closer to the observer, to the south in our maps) peak at each velocity, distance from the disk center to 
principal peaks, half power size of the emission distribution, and velocity at which 
the maximum intensity is present. These represent a total of 43 different 
observational parameters for each input model. 
\vspace{0.5cm}

\subsection{Line emission maps}
In all maps obtained from our set of models we observe the same tendencies as \citet{Gom00}. 
Summarizing, we observe an asymmetry on both sides of the major axis, as
expected from optically thick emission. 
The areas further away from the observer (positive declination 
in Fig. \ref{fig:PPD_disk}) show line emission of higher intensity because the line of
sight intercepts areas of the disk closer to the central star, where
the gas is warmer. This result is confirmed in DM Tau spectral line
observations by \citet{Dar03}, where an inclination angle $ i\simeq
-37\degr$ was assumed. 
We notice another asymmetry on both sides of the minor axis, that it is more pronounced when approaching to the 
velocity of the cloud ($v=0$ km s$^{-1}$). It is caused by the asymmetry of the hyperfine structure in the 
C$^{17}$O molecular transitions.

We find that the maximum intensity of the line at the systemic velocity 
($v=0$ km s$^{-1}$) traces the outer edge of the disk.
As previously discussed by  \citet{Sar91} and
\citet{Gom00}, this is a consequence 
of the fact that the effective area emitting at a given 
velocity within the beam increases with the distance to the central star
more steeply than the decrease of brightness temperature with distance.

Moreover, at the center of the disk the emission intensity diminishes, and 
we can even see absorption lines in some of our maps. The high dust continuum
opacities typical in the central parts of the disk, reduces the contrast between
the emission lines and the continuum, and, if it is high enough, lines
could show up in absorption.

\section{Statistical tools}
\label{PPD:sta}

As we mentioned in section~\ref{PPD:gri}, we chose a set
of 43 different observational parameters to characterize the maps
resulting from our model calculations. These parameters are, in
principle, somewhat arbitrary. If we want to simulate a real
observation, in which we would like to extract information about the
underlying physical characteristics of disks, it is obvious that some
of these 43 parameters will have more informative power, while some
may turn out to be irrelevant. Moreover, it is likely that not all the
chosen observational parameters will be independent.

Here, we have undertook a statistical analysis to try to identify a
set or a combination of observational parameters that could render
more information about the disks properties. First, we used a
principal component analysis to reduce the number of observational
parameters to a small, informative set. Later, we investigate whether
we can obtain quantitative values of physical magnitudes from
observational parameters, by means of multiple linear regression.

\subsection{Principal Components}

The principal component analysis is a statistical technique that provides
a dimensional reduction of a set of variables. 
In our case, each of the initial 43 observational parameters would be
an axis in a system of coordinates in a multidimensional space. The method consists
of finding a set of orthogonal axes in which the variance 
(heterogeneity) of our data is maximum (see \citealt{Thu47,Kai58}). 
This is solved through a linear and orthogonal transformation that
corresponds to a rigid rotation of the original data into a new set of
coordinates. 
The eigenvalues obtained provide information about the variances of
the data in the new space, and the eigenvectors represent the
direction of the axes in the new space of representation of our data.   
In our study, this analysis reduces the number of observational signatures necessary to 
derive information about the physical parameters, to a smaller set of linearly
 independent parameters: the principal components.

\subsection{Multiple linear correlation}

In order to quantitatively estimate each physical parameter from the set of 
observational variables, we have also carried out a multiple regression analysis \citep{Pea08}.
In our case, the obtained principal components will be considered the
independent variables, and the physical properties of the disk will be
the dependent ones. The multiple linear correlation analysis will then
try to make the best fit, to derive a
linear function of the form:
$$y_i = a_1 x_1 + a_2 x_2 + ... + a_j x_j \;,$$
where $y_i$ is the dependent variable, and $x_j$ are the independent
ones. In an ideal case (correlation coefficient $\simeq 1$), we would
obtain a function with a good predictive power for the dependent
variable. For our particular problem, we would like to obtain a formula
with which, from a set of observational parameters, we could calculate
the physical characteristics of the disk.

Moreover, to test the signification of the multiple regression we applied the 
F-Snedecor test \citep{Sne34} as hypothesis testing. The ratio of two chi-squares divided by their 
respective degrees of freedom follows an F distribution. The test consist of comparing the 
relation between the variance of the predicted values for the
dependent valuable and the error variance with the value of the
 F distribution.

\section{Observational signatures vs. physical parameters}
\label{PPD:res}

\subsection{Eigenvalues and Eigenvectors of Principal Components} 
To reduce the number of relevant components in the principal component
analysis, we adopted the  Kaiser
criterion \citep{Kai60}, which only retains
factors with  eigenvalues greater than 1. In our case, the criterion selects
four factors, which account for 91\% of the total variance of the 
system.  In fact, the first and the second factors alone  represent
 82\% of the total variance. Therefore, most of the following analysis is based on
 these  two first principal components.

We have represented the eigenvectors in Table \ref{tb:PPD_eigenvectors}, for each principal component.
The numerical entries in this table show the linear combination
coefficients for each observational parameter, used to build the
corresponding principal component. Therefore, they indicate the
relative weight of each observational parameter on the components.

\begin{table}[!h]
\setlength{\tabnotewidth}{\columnwidth}
  \tablecols{5}
{\scriptsize
\begin{center}
\caption[Eigenvectors]
{Eigenvectors}
\label{tb:PPD_eigenvectors}
\vspace{2mm}
  \begin{tabular}{lrrrr}
    \toprule \toprule
\multicolumn{1}{c}{Observational}&
\multicolumn{1}{c}{PC1}&
\multicolumn{1}{c}{PC2}&
\multicolumn{1}{c}{PC3}&
\multicolumn{1}{c}{PC4}\\
\multicolumn{1}{c}{Parameters}& &&&
\\
\midrule
I(v$_{0}$)\tabnotemark{a}           & -0.0279       &0.0020           &0.0030          & 0.0033\\
I(v$_{0.5}$)\tabnotemark{a}            & -0.0276       &0.0019           &0.0029          &0.0034\\
I(v$_{1}$)\tabnotemark{a}            & -0.0281       &0.0007           &0.0022          &0.0035\\
I(v$_{1.5}$)\tabnotemark{a}            &-0.0340        &-0.0023          &0.0024          &0.0040\\
I(v$_{2}$)\tabnotemark{a}            &-0.0276        &-0.0086          &0.0016          &0.0040\\
I(v$_{2.5}$)\tabnotemark{a}            &-0.0118        &-0.0082          &0.0028          &0.0040\\
r(v$_{0}$)\tabnotemark{b}            &-0.1360        &0.0388           &-0.0072         &-0.0137\\
r(v$_{0.5}$)\tabnotemark{b}            &-0.1462        &0.0325           &-0.0041         &-0.0169\\
r(v$_{1}$)\tabnotemark{b}            &\textbf{-0.1581}\tabnotemark{*} &0.0448           &-0.0015         &-0.0086\\
r(v$_{1.5}$)\tabnotemark{b}          &\textbf{-0.1930} &0.0100           &0.0266          &-0.0262\\
r(v$_{2}$)\tabnotemark{b}            &-0.1077        &0.0075           &-0.0030         &-0.0150\\
r(v$_{2.5}$)\tabnotemark{b}            &-0.0483        & 0.0310          &-0.0094         &-0.0062\\
I$_{sec}$(v$_{0}$)\tabnotemark{c}   &-0.0289        &-0.0047          &0.0004          &0.0019\\
I$_{sec}$(v$_{0.5}$)\tabnotemark{c}   &-0.0297        &-0.0042          &0.0006          &0.0022\\
I$_{sec}$(v$_{1}$)\tabnotemark{c}   &-0.0179        &-0.0091          &-0.0024         &-0.0018\\
I$_{sec}$(v$_{1.5}$)\tabnotemark{c}   &-0.0003        &0.0007           &-0.0009         &-0.0004\\
I$_{sec}$(v$_{2}$)\tabnotemark{c}   &0.0000         &0.0000           & 0.0000         &0.0000\\
I$_{sec}$(v$_{2.5}$)\tabnotemark{c}   &0.0000         &0.0000           &0.0000          &0.0000\\
a(v$_{0}$)\tabnotemark{d}            &-0.0595        &0.0426           & 0.0316         &0.0034\\
b(v$_{0}$) \tabnotemark{d}           &-0.0790        &0.0334           &0.0272          &-0.0184\\
a(v$_{0.5}$)\tabnotemark{d}        &\textbf{-0.1615} &-0.0309          &0.0259          &0.0043\\
b(v$_{0.5}$)\tabnotemark{d}        &\textbf{-0.1947} &-0.0788          &\textbf{0.0523} &-0.0076\\
a(v$_{1}$)\tabnotemark{d}            &-0.1138        &-0.0250          &0.0281          &-0.0024\\
b(v$_{1}$)\tabnotemark{d}            &-0.0388        & -0.0024         &0.0041          &0.0017\\
a(v$_{1.5}$)\tabnotemark{d}        &\textbf{-0.1777} &\textbf{-0.1926} &-0.0350         &\textbf{-0.0615}\\
b(v$_{1.5}$)\tabnotemark{d}        &\textbf{-0.2271} &\textbf{-0.1499} &0.0086          &\textbf{0.0743}\\
a(v$_{2}$)\tabnotemark{d}          &\textbf{-0.2420} &-0.0171          &\textbf{0.0543} &-0.0023\\
b(v$_{2}$)\tabnotemark{d}            &-0.1181        &0.0056           &0.0133          &-0.0084\\
a(v$_{2.5}$)\tabnotemark{d}            &-0.0286        &0.0168           &-0.0054         &0.0011\\
b(v$_{2.5}$)\tabnotemark{d}            &-0.0028        &0.0114           &-0.0067         &0.0010\\
a$_{sec}$(v$_{0}$)\tabnotemark{e}     &-0.0746      &\textbf{0.1023}   &\textbf{0.0703} &\textbf{0.0434}\\
b$_{sec}$(v$_{0}$)\tabnotemark{e}    &-0.0436       &\textbf{0.1026}   &0.0406          &\textbf{-0.0607}\\
a$_{sec}$(v$_{0.5}$)\tabnotemark{e}    &-0.0026        & 0.0161          &-0.0127         &-0.0076\\
b$_{sec}$(v$_{0.5}$)\tabnotemark{e}    &-0.0006        &0.0047           &-0.0057         &-0.0011\\
a$_{sec}$(v$_{1}$)\tabnotemark{e}    &0.0000         &0.0000           &0.0000          &0.0000\\
b$_{sec}$(v$_{1}$)\tabnotemark{e}    &0.0000         &0.0000           &0.0000          &0.0000\\
a$_{sec}$(v$_{1.5}$)\tabnotemark{e}    &-0.0733    &\textbf{0.0985}    &\textbf{0.0683} &\textbf{0.0525}\\
b$_{sec}$(v$_{1.5}$)\tabnotemark{e}   &-0.0398     &\textbf{0.0951}    &0.0398          &\textbf{-0.0549}\\
a$_{sec}$(v$_{2}$)\tabnotemark{e}    &-0.0040        &0.0185           &-0.0141         &-0.0097\\
b$_{sec}$(v$_{2}$)\tabnotemark{e}    &-0.0008        &0.0043           &-0.0060         &-0.0006\\
a$_{sec}$(v$_{2.5}$)\tabnotemark{e}    &0.0000         &0.0000           &0.0000          &0.0000\\
b$_{sec}$(v$_{2.5}$)\tabnotemark{e}    &0.0000         &0.0000           &0.0000          &0.0000\\
v$_{I_{max}}$ \tabnotemark{f}      &\textbf{0.3176}  &\textbf{-0.1431} &\textbf{0.1527} &-0.0241\\ 
                                            
      \bottomrule \bottomrule

\tabnotetext{a}{Intensity of the principal peak at each velocity (0, 0.5, 1, 1.5, 2, 2.5 km s$^{-1}$) in Jy beam$^{-1}$.}
\tabnotetext{b}{Distance from disk center to principal peaks in arcsec.}
\tabnotetext{c}{Intensity of the secondary peak at each velocity.}
\tabnotetext{d}{Minor (a) and mayor (b) axis of the half power sizes of emission for the principal peak of intensity at each different velocity, in arcsec.}
\tabnotetext{e}{Minor (a) and mayor (b) axis of the half power sizes of emission for the secondary peak of intensity at each different velocity.}
\tabnotetext{f}{Velocity at which the maximum intensity is present in km s$^{-1}$.}
\tabnotetext{*}{Values in boldface represent the observational parameters with a larger weight in the definition of each PC.}
\end{tabular}
\end{center}
}
\end{table}


For the first principal component (PC1) the observational parameters with a
 larger weight in its definition are (in order of decreasing relative
 weights) the velocity of the peak emission, the half power sizes for 
principal peaks at intermediates velocities, and the distance from
 principal peaks to center. The parameters that define the second
 principal component (PC2) are the half power sizes of principal peaks
 at 1.5 km s$^{-1}$, the velocity of the peak emission, and the half
 power sizes of secondary peaks at 0.0 and 1.5 km s$^{-1}$.
The third principal component (PC3) is defined by the velocity of the
peak emission and the half power sizes for principal peaks at 0.5 and 2.0 km s$^{-1}$, and for secondary peaks at 0.0 and 1.5 km s$^{-1}$.
Finally, the physical parameters that define the forth  principal
component (PC4) are the half power sizes for principal peaks at 1.5  
km s$^{-1}$ and for secondary peaks at 0.0 and 1.5 km s$^{-1}$. 
We have marked in boldface these most representative observational parameters in Table~\ref{tb:PPD_eigenvectors}.

\subsection{PC1(kinematical component)-PC2(spatial component) diagrams}
\label{PPD:res_PC}
Due to the fact that the first and the second principal
component represent the 82\% of the total variance of the
system, we have represented all our models in a
PC1-PC2 diagram, to check whether such diagrams can be used to discriminate
disks with particular physical characteristics, if we see some clustering
related to the physical properties. Considering the parameters with
the larger weights in each of these principal components, we have named them
as  ``kinematical component'' and ``spatial component'' in the
case of PC1 and PC2, respectively.

\begin{figure}[!h]
\begin{center}
\includegraphics[width=\columnwidth]{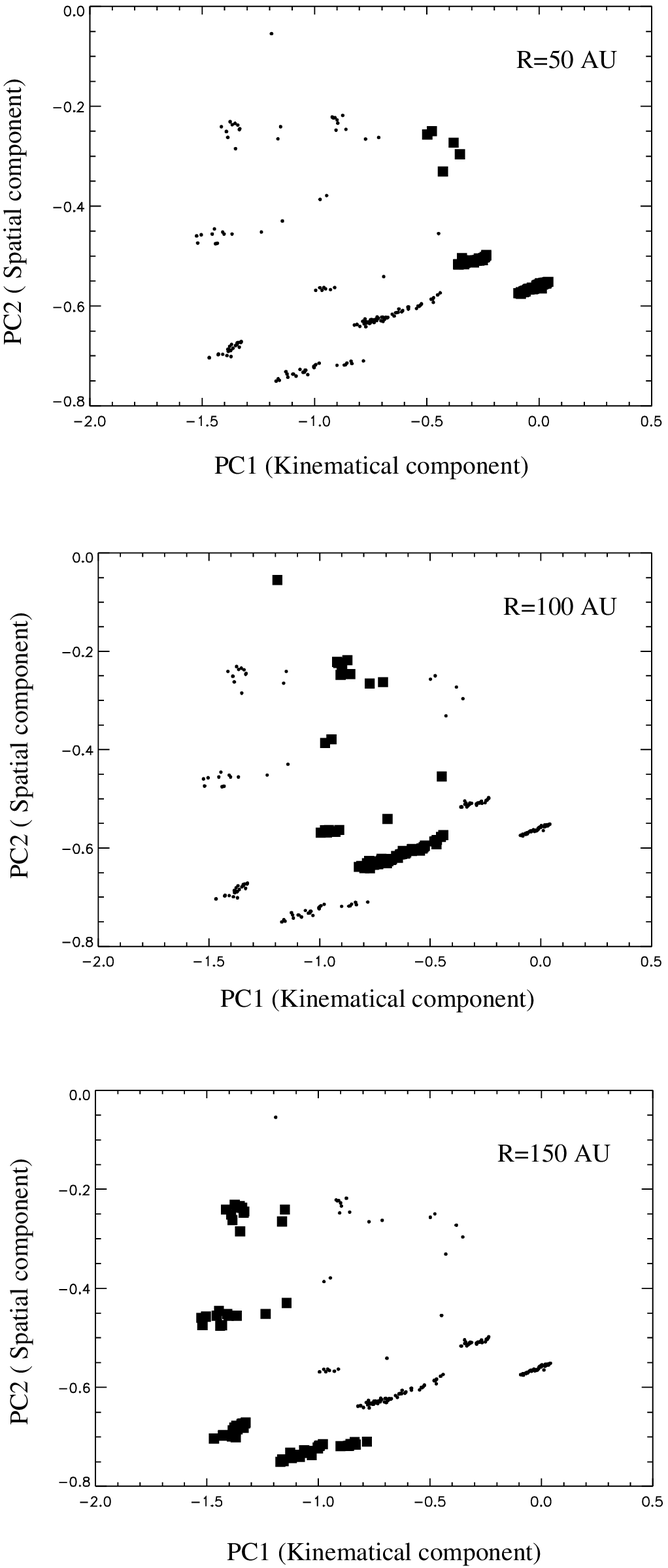}
\end{center}
\vspace{0.13cm}
\caption[PC1-PC2 diagrams: Radius]
{Disk radii in the PC1-PC2 diagrams. Dots represent all the modeled disks
  with different physical parameters. Squares represent disks with
  radius 50 A.U. (upper), 100 A.U. (central) and 150 A.U. (bottom).} 
\label{fig:PPD_PCR}
\vspace{0.4cm}
\end{figure}

The most evident result is shown when we represent  disks with
different radii in the PC1(kinematical)-PC2(spatial) 
diagram (see Fig \ref{fig:PPD_PCR}). The disks with radius 50 AU are
distributed on the right part of the diagram, the ones with radius 
100 AU are located at the central part and the disks with radius 150 AU are 
located on the left of the plot. This means that the kinematic 
principal component (x axis) is 
qualitatively good to discriminate among disks with different radii.  
This result indicates  that we can get information about the radius of a protoplanetary disk  
(physical parameter) from the velocity of the peak emission, the half power sizes 
for principal peaks at intermediates velocities and from the distance from principal
peaks to the center 
(observational signatures). That the disk radii can be discriminated
relatively well by maps of line emission may seem a relatively obvious result, but
it illustrates the power of this kind of statistical analysis.

\begin{figure}[!h]
\begin{center}
\includegraphics[width=\columnwidth]{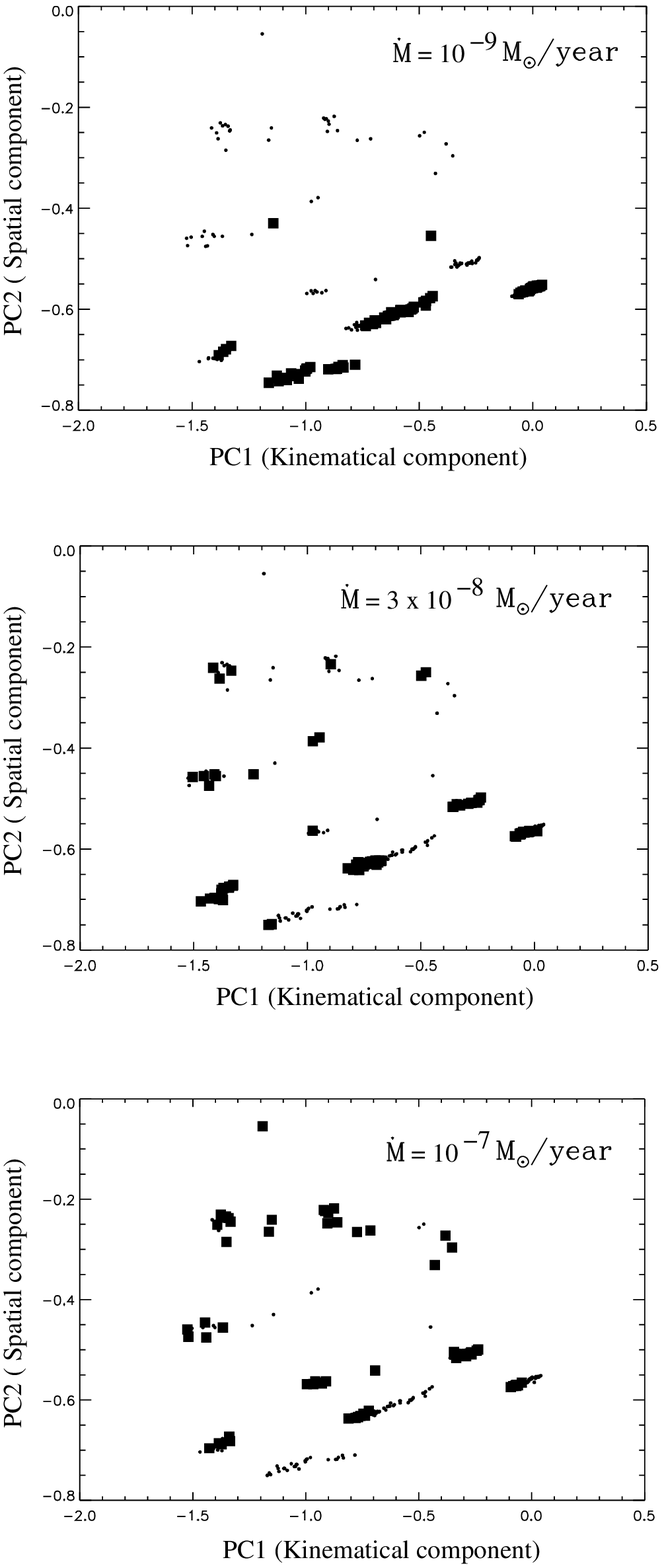}
\end{center}
\caption[PC1-PC2 diagrams: $\dot{M}$]
{Mass accretion rates in the PC1-PC2 diagrams. Dots represent all the modeled disks
  with different physical parameters. Squares represent disk with mass accretion rate of 10$^{-9}$  M$_{\odot}$/year (upper), 3$\times$10$^{-8}$  M$_{\odot}$/year (central) and 10$^{-7}$  M$_{\odot}$/year (bottom).}
\label{fig:PPD_PCMp}
\end{figure}

Another interesting trend is seen from the representation in the diagram of disks with 
different mass accretion rate (see Fig \ref{fig:PPD_PCMp}). Disks with higher mass
accretion rates tend to populate the upper parts of the PC1(kinematical)-PC2(spatial) diagram.
Therefore, in this case it is the second principal component (y axis) the one
that better discriminates among disks with 
different mass accretion rates. Considering the parameters that give rise to
this component, we can say that the half power sizes of principal peaks at 1.5 km s$^{-1}$, 
the velocity of the peak emission and the half power sizes of secondary peaks at 0.0 and 1.5 km s$^{-1}$ 
velocities provide information about the mass accretion rate of the disks. 

Other trends relating principal components and the rest of the physical parameters are also present, 
but qualitatively they are not as clear as the two we have mentioned. In
the case of other physical parameters, it is the second 
principal component the one that provides more information about the $\alpha$
parameter and the maximum radius of dust grains. Further analysis including models of more
molecular transitions will certainly be useful to obtain refined principal
components that can better discriminate these physical parameters.

\subsection{Multiple correlation}

We have seen in the previous section that diagrams of principal components can
be useful to discriminate among disks with different physical
characteristics. It would  be interesting, however, to have a mathematical
tool to easily obtain numerical values for these physical
characteristics based on the observed maps.
As a first approach, we have here performed a multiple linear
correlation with the observational parameters as independent
variables, and the physical characteristics as dependent ones. 
The resulting coefficients for such a fit are shown in 
Table~\ref{tb:PPD_correl}.

As it was deduced in the previous
subsection, the linear correlation coefficients of each principal component 
(see r(PC$_{i}$) values in Table~\ref{tb:PPD_correl}) show that PC1(kinematical component) is the
best one to derive
information about the disk radii and PC2(spatial component) provide the
most information in the determination of the rest of the physical
parameters, $\dot{M}$, $\alpha$, and maximum radius of dust grains. 

\begin{table*}[!t]\centering
\setlength{\tabnotewidth}{0.9\textwidth}
  \tablecols{7}
{\small
\caption[Correlation coefficients and F-Snedecor test
  values]{Correlation coefficients and F-Snedecor test
  values}
\label{tb:PPD_correl}
\vspace{2mm}
  \begin{tabular}{ccccccc}
    \toprule \toprule
&
\multicolumn{1}{c}{r(PC1)\tabnotemark{a}}&
\multicolumn{1}{c}{r(PC2)\tabnotemark{a}}&
\multicolumn{1}{c}{r(PC3)\tabnotemark{a}} &
\multicolumn{1}{c}{r(PC4)\tabnotemark{a}}&
\multicolumn{1}{c}{R\tabnotemark{b}}&
\multicolumn{1}{c}{F}
\\
\midrule
Radius            &   -0.9&    -0.19&  0.07& -0.001&  0.97&  951.62 \\
$\dot{M}$         &   -0.18&    0.5&  -0.3&  -0.16&   0.57&   26.93  \\
$\alpha$          &    0.16&   -0.21&  0.13&  0.07&   0.31&    5.94  \\
Max. Grain Radius &   -0.08&   -0.009& -0.012& 0.08&  0.19&    2.07  \\
      \toprule \toprule
 &            $\lambda_{0}$\tabnotemark{c} &$\lambda_{1}$\tabnotemark{c}  &$\lambda_{2}$\tabnotemark{c}  &$\lambda_{3}$\tabnotemark{c}  &$\lambda_{4}$\tabnotemark{c} \\       
\midrule
Radius            & -55.3&  -79.4&  -54.2& 133.3& -137.7 &\\          
$\dot{M}$         &  1.8$\times$10$^{-7}$&  -0.16$\times$$10^{-7}$& 1.4$\times$10$^{-7}$& -1.7$\times$10$^{-7}$& -0.15$\times$10$^{-7}$ &\\
$\alpha$          & -0.0195& 0.0024& -0.0349& +0.0465& -0.0091 &\\              
Max. Grain Radius & -16943.4& -100.4& -40213.2& +48703.1& +123751.8 &\\
\bottomrule \bottomrule

\tabnotetext{a}{r(PCi) are the linear correlation coefficients of
  each principal component.}
\tabnotetext{b}{Multiple correlation coefficient.}
\tabnotetext{c}{$\lambda_{i}$ are the coefficients of the linear
  combination of principal components obtained from the multiple
  regression study, to derive the values of the physical
  parameters in the first column.}
\end{tabular}
}
\end{table*}

The most important piece of information that can be retrieved with
this kind of study is the deduction of the linear combination of observational
signatures that provide quantitative information about physical characteristics.
This linear combination can be derived from $\lambda_{i}$ coefficients in Table~\ref{tb:PPD_correl} as follows:
$$P~= \lambda_{0}~ +~\lambda_{1}PC1~+~\lambda_{2}PC2~ +~\lambda_{3}PC3~ +~\lambda_{4}PC4$$
where P is the physical parameter.
As an example, in the case of the radius of the protoplanetary disk, this
combination would be the following:

 $$R_d = -55.3-(79.4\times PC1)-(54.2\times PC2)+$$ 
$$  +(133.3 \times PC3)-(137.7\times PC4)$$

With these sets of linear combinations, we could estimate physical
parameters from observations, provided that the fit is good enough.

The strongest correlation (see R values in Table~\ref{tb:PPD_correl}) is obtained for radius 
(correlation coefficient R = 0.97), followed by mass accretion rate
(R = 0.57), viscosity parameter (R = 0.31) and maximum radius of dust grains
(R = 0.19). 
The multiple correlation coefficients alone are not good statistical indicators
of the goodness of linear fits.
To assess the validity of our fitted functions we carried out an F test to
derive the signification of the linear
regression. We assume a confident limit of 95\%, which means that we
could admit as good fits with values $F < 1.44$. However in all our results
the values exceeded this critical value (see Table \ref{tb:PPD_correl}). This
result suggests that our variables are far from the linear regime,
which is certainly reasonable.

\subsection{Effects of  measurements errors}

In our method, the derived value of a physical parameter depends linearly on the principal components, which are linear combinations of the observational data. Therefore we must consider on one hand the error in the calculation of the principal components, and on the other hand the error derived from the calculation of the multiple correlation. 

In a real observation, the error of the intensity will depend on the rms noise reached during the observations, and, in particular, the achieved signal-to-noise ratio will determine the positional accuracies (1$\sigma$ error in position $\simeq HPBW/(2\;{\rm SNR})$), assuming a compact emitting source). In our method, for the determination of the most significant principal components, the most important sources of error are related to the observational parameters that contribute with a larger weight, i.e., the half power sizes of emission for the principal and
secondary peaks of intensity (whose error can be roughly estimated as
$\simeq HPBW/{\rm SNR}$), and the velocity at which the maximum intensity is present (with 
$1\sigma$ error $\sim\Delta v/({\rm 2\; SNR})$, where $\Delta v$ is the linewidth). 
Since principal components are linear combinations of observational parameters, the weight of the errors associated with each measured observational parameters will depend on theirs particular weight in the definition of each principal component (see Table~\ref{tb:PPD_eigenvectors}). 
Nevertheless, we must point out that, in the case of calculations of the physical parameters of
disks, the error is dominated by the uncertainties in the fits of the 
multiple regression. Since our derived variables are far from linear regime (see F test above),measurements errors are not significant in comparison.

\section{Comments and prospects for these statistical studies}
\label{PPD:com}

The results derived from the statistical study presented in this
paper show that this method could be a powerful tool to obtain
information related to the physical characteristics in protoplanetary
disks from observational parameters. 

The most important information of our study is related to
the radius of the disks. 
This preliminary study results in a way to obtain a first approximation of the disks radii that depends 
only on the observations, avoiding the application of $\chi^2$ fitting
techniques.

Since this is a promising method of study, and considering the future
set of observations in protoplanetary disks that will be carried out
with the development of ALMA, we plan to complete and improve our
models and analysis.

 Nowadays it is not possible to test observationally our method
 because there is no observation of protoplanetary disks in our selected molecular 
transition, C$^{17}$O(J=3$\rightarrow$2). But, considering the excellent characteristics of this transition/isotope to observe protoplanetary disks (see subsection~\ref{PPD:rad_line}), one should be able to test our study in the near future with ALMA observations.
Meanwhile this moment comes, we will improve or study. 

First, to better determine the tendencies in the principal component diagram, we plan to
increase the sample, building a more extended set of models with a wider
variety of initial physical and observational parameters. Calculations in other
molecular transitions or isotopes would further
constrain the information on temperature and 
density. An update in the calculation of the disk structure
models would be done, considering in detail the photodisociation or
depletion effects in the line of study, along with dust grain growth and settling. 
We also plan to study the application of our method towards non-axial-symmetric disks, 
for what the development of a new disk structure that takes into account the lack of symmetry is needed.
Finally, the determination of a more complex method to derive the physical parameters of disks from the principal components is desirable, since the application of multiple correlation provides variables far form the linear regime and, introducing the greater source of error in our study.

\section{Detectability of molecular lines with new interferometers}
\label{PPD:dec}

It would obviously be important to be able to test our results observationally in the
future. With this mind, we have also studied the detectability of the calculated
models, when observed with interferometers that will be able to reach subarcsecond resolution at the
frequency of the C$^{17}$O(J=$3\rightarrow 2$) transition, i.e., SMA and ALMA.

We have calculated the line intensity expected for all our line
models. The lowest peak intensity 
($\sim$7 mJy beam$^{-1}$) corresponds to disks with $R_{d}$=50 AU,
$\dot{M}$ = 10$^{-9}$ M$\sun$ yr$^{-1}$, $\alpha$ = 0.05 and 1 $\mu$m
maximum grain radius. On the other hand, the highest peak intensity ($\sim$180 mJy beam$^{-1}$)
corresponds to disks with 150 AU radius,
$\dot{M}$ = 10$^{-7}$ M$\sun$ yr$^{-1}$, $\alpha$ = 0.01 and 10$^{5}$ $\mu$m
maximum grain radius.

With the SMA, a sensitivity 
of $\sim$85 mJy beam$^{-1}$ is
expected\footnote{http://sma1.sma.hawaii.edu/call.html}, 
considering the 8 antennas of
the array working at 337 GHz, with 1 km s$^{-1}$
 velocity resolution and 10 hours integration time, under standard
 values of precipitable water vapor ($\sim$2.0 mm for 300$-$355 GHz). 

On the other hand, in the case of ALMA, 
considering 64 antennas, 1 km s$^{-1}$  velocity resolution, only 1
hour integration time and 1.5 mm of precipitable water vapor (median
at the site over all hours and seasons), the sensitivity expected is
1.2 mJy beam$^{-1}$ \citep{But99}.

Considering 5$\sigma$ emission as detections, we conclude that all 
our disks modeled, even the faintest, will be detected with one hour integration time with
ALMA. Nevertheless, it will be extremely difficult of observe any of
our modeled disks with the SMA, at least the ones with the physical
characteristics showed in this work (see
Table~\ref{tb:PPD_phys-param}), since the SNR obtained for the modeled disks
with the highest intensities ($\sim$180 mJy beam$^{-1}$) in 10 hours
of integration time, would only be 2$\sigma$.

Therefore, ALMA will be a crucial instrument to observationally test the predictions
and assumptions of our models.

\section{Conclusions}
\label{PPD:con}

In this paper we have presented a statistical method to derive physical
parameters from observational characteristics in a protoplanetary
disk. To make this study we modeled the expected emission of the
 C$^{17}$O(J=$3\rightarrow 2$) transition from
protoplanetary disks with different physical properties. We then
applied a principal component and a multiple correlation analysis, to
obtain a set of  linear combinations of observational parameters that
may 
provide relevant information of the physics of the disks. The main conclusions
are the following:

\begin{itemize}
\item The most significant results of our analysis are related to  disk sizes. We can
discriminate among disks with different
radii using a principal component composed mainly of the velocity of the peak emission, the 
half power sizes for principal peaks at intermediates velocities and the distance from
 principal peaks to center. 

\item Moreover,  some information about the mass accretion rate could
  be obtained from a principal component made of the half
power sizes of principal peaks at 1.5 km s$^{-1}$, the velocity of 
the peak emission and the half power sizes of secondary peaks at 0.0 and 1.5 km s$^{-1}$ 
velocities, although the results are much less significant than in the
case of radii.

\item Although with preliminary results, the statistical 
  method presented here seems to be promising and useful, and will be 
  improved and completed in a future with studies in other transitions.

\item  We have performed a study of detectability with ALMA and SMA. All our
modeled disks could be detected with ALMA, using one hour integration
time, nevertheless the sensitivity reached by the SMA is not enough
to detect our disks with reasonable integration times. We conclude that ALMA will play an important role
to test observationally our models and our statistical results. 

\end{itemize}

\acknowledgements

We wish to thank the LAEFF staff for allowing the use of their personal computers to carry out the calculations of the molecular line emission models.
PD acknowledges grants from CONACyT and PAPIIT, M\'exico.
The model calculations were performed in the cluster at CRyA-UNAM
acquired with CONACYT grants 36571-E, 47366-F, and UNAM-PAPIIT grant 110606.
FG and IdG are supported by Junta de Andaluc\'{\i}a grant FQM-1747. JFG 
also acknowledges partial support from grants AYA2005-08523-C03-03 of 
the Spanish MEC (including FEDER funds) and TIC-126 of Junta de 
Andaluc\'{\i}a.

\end{document}